\documentclass[aps,prd,onecolumn,preprintnumbers,nofootinbib,noshowpacs,superscriptaddress,letterpaper,floatfix,10pt]{revtex4-2}

\pdfoutput=1
\usepackage{graphicx}

\usepackage{amsmath,amssymb}
\usepackage{url}
\usepackage{multirow}
\usepackage[normalem]{ulem}
\usepackage{slashed}
\usepackage{rotating}
\usepackage{afterpage}
\usepackage{ifthen}
\usepackage{mciteplus}
\usepackage{multirow}
\usepackage{color}
\usepackage{pifont}

\newcommand{\be}{\begin{equation}}
\newcommand{\ee}{\end{equation}}
\newcommand{\bea}{\begin{eqnarray}}
\newcommand{\eea}{\end{eqnarray}}
\newcommand{\ba}{\begin{array}}
\newcommand{\ea}{\end{array}}

\usepackage[normalem]{ulem}
\usepackage{xcolor}

\definecolor{grey}{rgb}{0.5,0.5,0.5}

\usepackage[
    colorlinks=true,
    linkcolor=blue,
    citecolor=blue,
    urlcolor=blue,
    breaklinks=true,
    bookmarks=true
]{hyperref}

\begin{document}

\title{Neutrino backgrounds in matter-wave interferometry: implications for dark matter searches and beyond-Standard Model physics}

\author{Jo\~ao Paulo Pinheiro}
\email{joaopaulo.pinheiro@fqa.ub.edu}
\affiliation{Departament de Física Quàntica i Astrofísica and Institut de Ciències del Cosmos, \\
  Universitat de Barcelona, Diagonal 647, E-08028 Barcelona, Spain}


\begin{abstract}
We present a comprehensive theoretical analysis of neutrino-induced decoherence 
in macroscopic matter-wave interferometry experiments designed to search for dark matter and beyond-Standard Model 
physics. Our calculation includes contributions from the cosmic neutrino background (C$\nu$B), 
solar neutrinos, and reactor antineutrinos, accounting for coherent scattering processes across nuclear, 
atomic, and macroscopic length scales. Within the Standard Model, we find negligible decoherence rates for 
planned experiments such as MAQRO ($s/\sigma_s \sim 10^{-27}$) and terrestrial interferometers like 
Pino ($s/\sigma_s \sim 10^{-22}$). However, these experiments achieve competitive sensitivity to 
beyond-Standard Model physics through light vector mediator interactions, with C$\nu$B constraining 
coupling products to $g_\nu g_n \lesssim 10^{-17}$ for $Z'$ masses below 1 eV. 
Our results provide a theoretical framework for interpreting matter-wave interferometry measurements in terms of neutrino interaction physics and for deriving constraints on BSM models from experimental data.

\end{abstract}

\maketitle
\section{Introduction}
\label{sec:intro}
The search for dark matter through direct detection has achieved extraordinary sensitivity to nuclear recoils above keV energies~\cite{Schumann:2019eaa}, yet fundamental limitations emerge when probing ultra-light dark matter candidates with masses below the GeV scale~\cite{Arvanitaki:2016fyj,Graham:2016plp,Geraci:2016fva,Graham:2017pmn,Stadnik:2015kia}. These limitations arise from detector energy thresholds that render soft scattering events unobservable, despite potentially large interaction cross sections. Matter-wave interferometry has emerged as a complementary approach, offering threshold-free sensitivity to arbitrarily soft interactions through collisional decoherence mechanisms~\cite{Du:2022ceh,Riedel:2016acj}.

The theoretical foundation exploits a fundamental quantum mechanical principle: any scattering process that transfers momentum along an interferometer baseline can extract which-path information and destroy spatial coherence~\cite{Zurek:2003zz,Hornberger:2006xxx}. Unlike conventional direct detection, which requires measurable energy depositions, interferometric detection relies solely on momentum transfer components that can be arbitrarily small. This enables access to kinematic regimes where $|\mathbf{q}| \sim 1/|\Delta x| \sim 10^7$ m$^{-1}$ for typical baseline separations $|\Delta x| \sim 100$ nm.

Coherent scattering enhancements are particularly essential to provide exceptional sensitivity for matter interferometers. When the de Broglie wavelength of probe particles exceeds the target size ($\lambda_{\rm probe} \gtrsim r_{\rm target}$), scattering amplitudes from individual constituents add coherently, yielding cross sections that scale as $N^2$ rather than $N$ for systems with $N$ scatterers~\cite{Riedel:2012ur,Riedel:2016acj,Du:2022ceh}. However, recent analyses by Badurina, Murgui, and Plestid~\cite{Badurina:2024ccd,Murgui:2025unt} have revealed important subtleties: while coherent enhancements always appear at the level of the $N$-particle density matrix, their observability depends critically on initial state preparation and measurement strategy.

This distinction proves crucial for different experimental configurations. In atom interferometers, individual atoms are prepared in uncorrelated superposition states and measured via one-body observables, which do not exhibit enhanced decoherence rates. In contrast, matter interferometers employ solid objects like nanoparticles containing $N \sim 10^{10}$ atoms whose relative positions are rigidly correlated through inter-atomic forces. When placed in spatial superposition, these systems realize highly entangled many-body states analogous to $N00N$ states, yielding full $N^2$ enhancement in decoherence rates for probe particle interactions~\cite{Badurina:2024ccd,Murgui:2025unt}.

Meanwhile, experiments are advancing rapidly, with demonstrations extending to increasingly massive objects~\cite{Arndt:1999kyb,Gerlich:2011wr,Eibenberger:2013dva}. Ambitious proposals now target space-based interferometry~\cite{Bateman:2014ena,Kaltenbaek:2015kha}, while large-scale terrestrial projects advance worldwide~\cite{Badurina:2019hst,Abe:2021ksx,Canuel:2018bzt,Zhan:2019quq}. These developments position matter-wave interferometry as a highly competitive probe complementary to traditional direct detection methods~\cite{Billard:2021uyg}.

However, this sensitivity presents a critical interpretational challenge: environmental backgrounds from Standard Model sources must be precisely characterized to distinguish genuine beyond-Standard Model signals. Previous analyses have addressed electromagnetic~\cite{Hornberger_2012,Stickler:2021eua}, gravitational~\cite{Bahrami:2014gwa,Marletto:2017wuj}, and gaseous collision backgrounds~\cite{Hackermuller:2003vf,Gallis:1990bd}, yet neutrino interactions remain largely unaddressed despite their fundamental importance.

This represents a critical gap for several compelling reasons. First, neutrino fluxes are omnipresent and cannot be shielded, with the cosmic neutrino background (C$\nu$B) providing approximately 112 neutrinos per cm$^3$ per species at energy scales $\mathcal{O}(\text{meV-eV})$ directly relevant to interferometer sensitivities~\cite{Vitagliano:2019yzm}. Second, the ultra-low energy regime naturally overlaps with neutrino scales, particularly for relic neutrinos whose non-relativistic velocities enable coherent interactions with macroscopic targets. Third, matter-wave interferometry could potentially achieve the first direct laboratory detection of the cosmic neutrino background.

Beyond their role as backgrounds, neutrino interactions provide novel discovery channels for BSM physics. Light vector mediators could significantly enhance interaction rates, either mimicking dark matter signatures or providing independent constraints.

In this work, we present the first systematic calculation of neutrino-induced decoherence in matter-wave interferometry experiments. Our analysis encompasses the complete spectrum of neutrino sources—cosmic neutrino background, solar neutrinos, and reactor antineutrinos—across interaction regimes spanning nuclear, atomic, and macroscopic coherent scattering. We treat both Standard Model neutral current interactions and representative BSM scenarios involving light mediators and electromagnetic couplings.

We demonstrate that while SM neutrino backgrounds remain negligible for all proposed experiments, matter-wave interferometry achieves competitive sensitivity to BSM interactions, providing the theoretical foundation for interpreting experimental results as robust constraints on new physics.

\section{Neutrino induced decoherence rate}
\label{sec:decoherence_intro}

Matter-wave interferometry exploits the wave nature of massive particles to create macroscopic quantum superposition states. Following the formalism established in Ref.~\cite{Du:2022ceh}, we parameterize the accumulated decoherence through a complex amplitude factor:
\begin{equation}
  \gamma\equiv \exp(-s+i\phi),
\label{eq:gamma_def}
\end{equation}
where $s$ represents the contrast loss (decoherence) and $\phi$ denotes a coherent phase shift acquired during the interferometer operation.

For an interferometer with target mass $M_{\rm tgt}$ operating over measurement time $t_{\rm exp}$, the decoherence accumulates according to:
\begin{equation}
  \ln \gamma = - M_{\rm tgt}\int_0^{t_{\rm exp}} R(t) \, dt \simeq - M_{\rm tgt} t_{\rm exp} R,
\label{eq:decoherence_accumu}
\end{equation}
where $R$ is the interaction rate responsible for decoherence, and in the second equality, we assume a time-independent rate, which is an excellent approximation for all neutrino sources considered in this work. 

The decoherence rate $R$ depends dominantly on three factors: the incident neutrino flux, the interaction cross sections with different target components, and the geometric probability that a given scattering event contributes to decoherence:
\begin{equation}
R = \sum_{i} \int \Bigg( \frac{d\Phi_\nu}{dE_\nu}\Bigg)\,\, \Big(N_i \, d\sigma_{\nu,i}\Big) \,\, (p_{\rm decoh}) \,\, dE_\nu,
\label{eq:rate}
\end{equation}
where $d\Phi_\nu/dE_\nu$ is the differential neutrino flux, $N_i$ represents the number of scattering targets of type $i$ (electrons, nucleons, atoms, etc.), $d\sigma_{\nu,i}$ is the corresponding differential cross section, and $p_{\rm decoh}$ encodes the decoherence probability.

\subsection{Multi-scale coherent interactions}

Matter-wave interferometry targets are inherently composite systems with hierarchical structures spanning multiple characteristic length scales. The fundamental constituents—electrons and nucleons—organize themselves through electromagnetic and strong interactions into increasingly complex structures: nucleons bind to form nuclei ($r_{\rm nuc} \sim$ fm), electrons arrange in atomic orbitals around nuclei ($r_{\rm atom} \sim$ Å), and atoms assemble into macroscopic targets ($r_T \sim \mu$m). Neutrinos with de Broglie wavelength $\lambda_\nu = 2\pi/|\mathbf{p}_\nu|$ can interact coherently with any subsystem whose characteristic size $r_{\rm tgt}$ satisfies the coherence condition $|\mathbf{q}|r_{\rm tgt} \lesssim 1$, where $\mathbf{q}$ is the momentum transfer. When this condition is met, the neutrino scatters coherently off all constituents within the subsystem, leading to dramatic cross section enhancements that scale as the square of the number of coherent scatterers.

This multi-scale coherent enhancement is uniquely accessible to matter-wave interferometry due to its threshold-free sensitivity to ultra-low momentum transfers. While conventional neutrino detectors require minimum recoil energies and thus probe only high-momentum-transfer processes, interferometric measurements can access the coherent scattering regime where $|\mathbf{q}|r_T \ll 1$ for macroscopic targets.

\subsubsection{Interaction regime classification}

Based on the coherence condition $|\mathbf{q}|r \lesssim 1$, we classify neutrino-target interactions into three distinct regimes that emerge naturally from the hierarchical structure of matter:

\vspace{0.3cm}
\textbf{Regime 1: Incoherent scattering} ($|\mathbf{q}|r_{\rm particle} \gg 1$)

In this regime, the neutrino wavelength is much smaller than individual particle sizes, precluding coherent interactions. Neutrinos scatter independently off nucleons and electrons, with cross sections scaling linearly with the number of target particles. This represents the familiar regime of conventional neutrino physics, where individual particle interactions dominate.

\vspace{0.3cm}
\textbf{Regime 2: Coherent nuclear and atomic scattering} ($|\mathbf{q}|r_{\rm nuc,atom} \lesssim 1$)

When the momentum transfer becomes sufficiently small, the neutrino wavelength becomes comparable to nuclear ($r_{\rm nuc} \sim A^{1/3}$ fm) or atomic ($r_{\rm atom} \sim Z^{-1/3}$ Å) length scales. This enables coherent interactions with entire nuclei or atoms, yielding cross section enhancements proportional to $A^2$ (for nuclear coherence) or $Z^2$ (for atomic coherence), where $A$ is the atomic mass number and $Z$ is the atomic number.

\vspace{0.3cm}
\textbf{Regime 3: Macroscopic coherent scattering} ($|\mathbf{q}|r_T \lesssim 1$)

In the ultra-low momentum transfer limit, neutrinos can interact coherently with the entire macroscopic target. This yields the maximum possible enhancement, with cross sections scaling as $N_A^2$, where $N_A$ is the total number of atoms in the target. This regime is particularly relevant for cosmic neutrino background interactions, where the extremely low energies ($\mathcal{O}(\text{meV})$) naturally satisfy the macroscopic coherence condition.

\subsubsection{Flux integration and source characteristics}

The total interaction rate incorporating all three regimes is:
\begin{equation}
R = \sum_{i=1}^3 \int \frac{d^3\Phi_\nu}{dE_\nu d\cos\theta_\nu d\phi_\nu} N_i \, d\sigma_{\nu,i}(M_i,p_\nu) \, p_{\rm decoh} \, dE_\nu d\cos\theta_\nu d\phi_\nu,
\label{eq:R_full}
\end{equation}
where $M_i$ and $N_i$ represent the effective mass and number of scatterers in regime $i$, and $d\sigma_{\nu,i}(M_i,p_\nu)$ denotes the differential cross section appropriate for each interaction type. The decoherence probability $p_{\rm decoh}$ ensures that only scattering events with momentum transfer components along the interferometer baseline contribute to the observable signal.

The implementation of Eq.~\eqref{eq:R_full} depends critically on the angular distribution of the incident neutrino flux. For directional sources such as solar neutrinos or reactor antineutrinos, the flux arrives from a well-defined direction $(\theta_0, \phi_0)$, allowing the angular integrals to be evaluated trivially:
\begin{equation}
R = \sum_{i=1}^3 \int \frac{d\Phi_\nu}{dE_\nu} N_i \, d\sigma_{\nu,i}(M_i,p_\nu) \, p_{\rm decoh} \, dE_\nu.
\label{eq:R_directional}
\end{equation}

For the cosmic neutrino background, the quasi-isotropic distribution necessitates careful angular integration, including the small but measurable anisotropy induced by Earth's motion ($\beta_\oplus \sim 10^{-3}$) relative to the CMB rest frame. Section~\ref{sec:NU_SOURCES} presents the neutrino sources considered in this analysis, detailing their flux characteristics and angular distributions.

\subsubsection{Decoherence probability}

The fundamental principle underlying matter-wave interferometry is that only momentum transfers with components along the interferometer baseline $\boldsymbol{\Delta x}$ can resolve the spatial separation and thereby induce decoherence. The decoherence probability quantifies this geometric constraint:
\begin{equation}
p_{\rm decoh} = 1 - \exp(i\mathbf{q} \cdot \boldsymbol{\Delta x}) = 1 - \exp(i|\mathbf{q}||\boldsymbol{\Delta x}|\cos\theta_q),
\end{equation}
where $\theta_q$ is the angle between the momentum transfer $\mathbf{q}$ and the baseline $\boldsymbol{\Delta x}$. 

This expression exhibits two distinct asymptotic behaviors that reveal the physical nature of the decoherence process. For small momentum transfers satisfying $|\mathbf{q}||\boldsymbol{\Delta x}| \ll 1$, the probability reduces to $p_{\rm decoh} \approx (\mathbf{q} \cdot \boldsymbol{\Delta x})^2/2$, demonstrating the quadratic suppression of soft scattering events. Conversely, for large momentum transfers where $|\mathbf{q}||\boldsymbol{\Delta x}| \gg 1$, the probability saturates at unity, indicating that hard scattering events provide complete which-path information and maximal decoherence~\cite{Du:2022ceh}.

\subsubsection{Cross section formulation}

We consider a general interaction framework where neutrinos couple to fermions $f \in \{e,p,n\}$ with coupling strengths $g_f$, assuming identical Lorentz structure for all fundamental vertices. For coherent processes involving spin-averaged amplitudes and unpolarized targets, only vector couplings contribute to the cross section, as axial-vector contributions cancel when summed coherently over target polarizations.

The differential cross sections for each interaction regime incorporate both the coherent enhancement factors and appropriate form factor suppression to ensure smooth transitions between regimes and prevent double-counting.
\footnote{In general, correlations exist between the coherent and incoherent scattering regimes, but we neglect these effects in this work for simplicity (for a detailed discussion on coherent and incoherent scattering and the proper treatment of correlations between these regimes, see~\cite{Bednyakov:2018mjd}). Since the coherent contribution scales as $N^2$ and the incoherent contribution scales as $N$, the correlation effects between them should scale as $N^{3/2}$, introducing corrections to our predictions by this factor. However, given that we are not performing data fits but rather testing sensitivity, this approximation is reasonable. A more precise calculation would require properly accounting for these correlations. }

\vspace{0.3cm}

\textbf{Regime 1: Incoherent scattering}

In this regime, neutrinos interact independently with individual fermions within the target. For a target composed of $N_A$ atoms with atomic number $Z$ and mass number $A$, we have $N_p = N_A Z$ protons, $N_n = N_A(A-Z)$ neutrons, and $N_e = N_A Z$ electrons. The total cross section becomes:
\begin{align}
d\sigma_{\nu,1} &= N_A \left[g_p^2\frac{Z}{A} d\sigma_{\nu,\rm fund}(M_N,p_\nu) + g_n^2\frac{A-Z}{A} d\sigma_{\nu,\rm fund}(M_N,p_\nu) \right.\nonumber\\
&\left.\quad + g_e^2\frac{Z}{A} d\sigma_{\nu,\rm fund}(M_e,p_\nu)\right] \times (1-|F_{\rm min}(|\mathbf{q}|r_{\rm min})|^2),
\label{eq:sigma_incoh}
\end{align}
where $d\sigma_{\nu,\rm fund}(M,p_\nu)$ represents the fundamental neutrino-fermion cross section for unit coupling and target mass $M$. The suppression factor $(1-|F_{\rm min}|^2)$ prevents double-counting as the coherence condition is approached, with $r_{\rm min}$ being the nuclear radius for nucleon interactions or the atomic radius for electron interactions.

\vspace{0.3cm}
\textbf{Regime 2: Coherent nuclear and atomic scattering}

This regime encompasses two distinct coherent processes that occur at different momentum transfer scales:

\emph{Nuclear coherence} ($|\mathbf{q}|r_{\rm nuc} \lesssim 1$): Neutrinos scatter coherently off entire nuclei with effective coupling $(Zg_p + (A-Z)g_n)$. The nuclear structure is encoded in the form factor:
\begin{equation}
F_{\rm nuc}(|\mathbf{q}|r) = \frac{3j_1(|\mathbf{q}|r)}{|\mathbf{q}|r} \exp\left(-\frac{|\mathbf{q}|^2 s_p^2}{2}\right),
\end{equation}
where $j_1(x)$ is the first spherical Bessel function, $r_{\rm nuc} = r_0 A^{1/3}$ with $r_0 = 1.2$ fm is the nuclear radius, and $s_p \approx 0.9$ fm accounts for nuclear surface diffuseness.

\emph{Atomic coherence} ($|\mathbf{q}|r_{\rm atom} \lesssim 1$): At even lower momentum transfers, coherent interactions encompass the entire atom (nucleus plus electron cloud) with total effective coupling $(Z(g_e + g_p) + (A-Z)g_n)$. The electronic structure is modeled using a Gaussian form factor:
\begin{equation}
F_{\rm atom}(|\mathbf{q}|r) = \exp(-r^2|\mathbf{q}|^2/2),
\end{equation}
where atomic radii $r_{\rm atom}$ are determined from atomic physics calculations.

The complete Regime 2 cross section incorporates both contributions with appropriate transition factors:
\begin{align}
d\sigma_{\nu,2} &= N_A (Zg_p + (A-Z)g_n)^2 |F_{\rm nuc}(|\mathbf{q}|r_{\rm nuc})|^2 \nonumber\\
&\quad \times [1-|F_{\rm atom}(|\mathbf{q}|r_{\rm atom})|^2] \, d\sigma_{\nu,\rm fund}(M_{\rm nuc},p_\nu) \nonumber\\
&\quad + N_A (Z(g_e + g_p) + (A-Z)g_n)^2 |F_{\rm atom}(|\mathbf{q}|r_{\rm atom})|^2 \nonumber\\
&\quad \times [1-|F_T(|\mathbf{q}|r_T)|^2] \, d\sigma_{\nu,\rm fund}(M_{\rm atom},p_\nu).
\label{eq:sigma_coh_nuclear}
\end{align}

The suppression factors $[1-|F|^2]$ ensure smooth transitions to higher-order coherent regimes while preventing overcounting of scattering contributions.

\vspace{0.3cm}
\textbf{Regime 3: Macroscopic coherent scattering}

In the ultra-low momentum transfer limit ($|\mathbf{q}|r_T \lesssim 1$), the neutrino wavelength exceeds the target size, enabling coherent interaction with the entire macroscopic object. This yields the maximum possible enhancement:
\begin{equation}
d\sigma_{\nu,3} = N_A^2 [Z(g_e + g_p) + (A-Z)g_n]^2 |F_T(|\mathbf{q}|r_T)|^2 d\sigma_{\nu,\rm fund}(M_T,p_\nu),
\label{eq:sigma_macro}
\end{equation}
where the $N_A^2$ enhancement reflects coherent scattering off all $N_A$ atoms simultaneously. The macroscopic form factor assumes a uniform spherical mass distribution:
\begin{equation}
F_T(|\mathbf{q}|r) = \frac{3j_1(|\mathbf{q}|r)}{|\mathbf{q}|r}.
\end{equation}

This regime becomes particularly important for cosmic neutrino background interactions, where the ultra-low energies ($\mathcal{O}(\text{meV})$) naturally satisfy $|\mathbf{q}|r_T \ll 1$ for typical interferometer targets. In Section~\ref{sec:CS_KIN}, we present the cross section calculation, including the kinematic constraints and amplitude computation.

\section{Neutrino Sources and Fluxes}
\label{sec:NU_SOURCES}

Having established the theoretical framework for multi-scale coherent interactions across three distinct regimes, we now examine the neutrino sources that enable experimental access to each regime. The energy hierarchy naturally maps onto the interaction regime classification: cosmic neutrino background (C$\nu$B) neutrinos with energies $\mathcal{O}(\text{meV})$ predominantly access macroscopic coherence (Regime 3), solar and reactor neutrinos with energies $\mathcal{O}(\text{keV-MeV})$ probe nuclear and atomic coherence (Regime 2), while the highest energy components approach the incoherent scattering regime (Regime 1).

This energy-regime correspondence is not merely convenient but fundamental: the coherence condition $|\mathbf{q}|r \lesssim 1$ directly relates neutrino energy to the maximum length scale over which coherent interactions can occur. For elastic scattering, the typical momentum transfer scales as $|\mathbf{q}| \sim E_\nu^2/(Mv)$ where $M$ is the target mass and $v$ is the relative velocity. Thus, lower energy neutrinos naturally access larger coherence volumes, enabling the dramatic cross-section enhancements that make matter-wave interferometry sensitive to ultra-weak interactions.

\begin{figure}[htbp]
\centering
\includegraphics[width=0.8\textwidth]{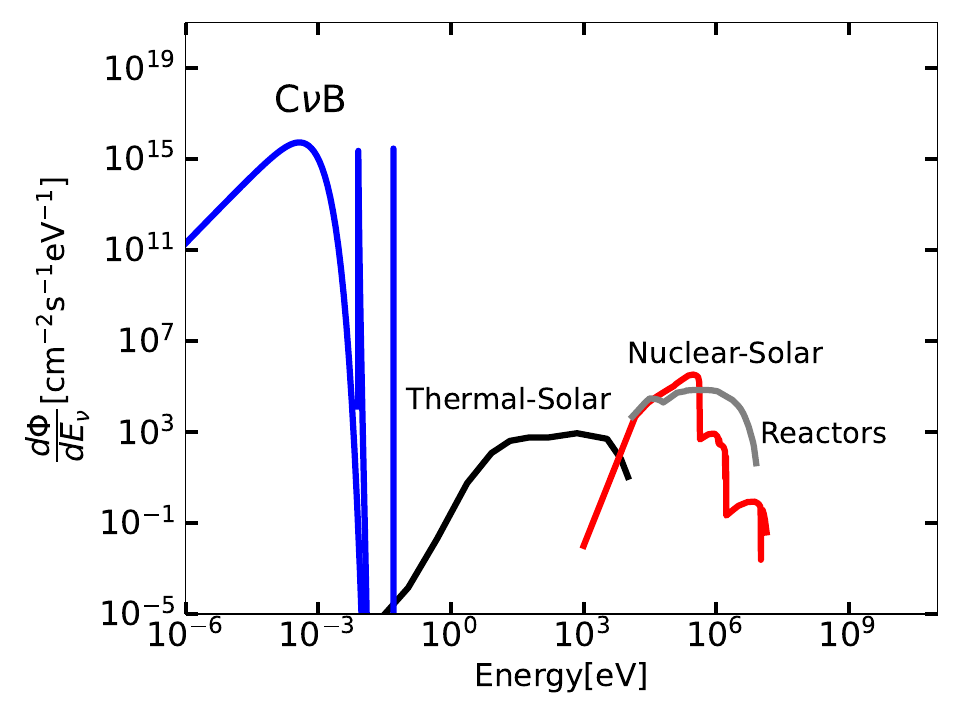}
\caption{Differential neutrino fluxes considered in this analysis spanning nearly eight orders of magnitude in energy. The cosmic neutrino background (C$\nu$B), represented in blue, shows contributions from three mass eigenstates: $m_1 = 0$ (continuous spectrum) and $m_2, m_3$ (sharp features at $E = m_\nu$). Solar neutrinos include both nuclear fusion (pp, CNO), represented in red, and thermal atmospheric components, represented in black. Reactor antineutrinos are normalized to 1 GW$_{\text{th}}$ at 100 m distance, represented in grey. Each source naturally accesses different coherent interaction regimes based on the energy-dependent coherence condition $|\mathbf{q}|r \lesssim 1$.}
\label{fig:fluxes}
\end{figure}

We consider three complementary neutrino sources that span nearly eight orders of magnitude in energy,  providing comprehensive coverage of the multi-scale coherent interaction framework. These fluxes are represented in FIG.~\ref{fig:fluxes}. The figure shows the dependence on the energy of the differential neutrino fluxes for each individual neutrino source. The cosmic neutrino background (C$\nu$B), represented in blue, shows contributions from three mass eigenstates: $m_1 = 0$ (continuous spectrum) and $m_2, m_3$ (sharp features at $E = m_\nu$). Solar neutrinos include both nuclear fusion (pp, CNO), represented in red, and thermal atmospheric components, represented in black. Reactor antineutrinos are normalized to 1 GW$_{\text{th}}$ at a distance of 100 m, represented in gray. Each source naturally accesses different coherent interaction regimes based on the energy-dependent coherence condition $|\mathbf{q}|r \lesssim 1$.

\subsection{Cosmic Neutrino Background}

The cosmic neutrino background represents the largest neutrino flux at Earth, consisting of relic neutrinos produced in the early universe. These neutrinos have a number density of approximately 112 cm$^{-3}$ per species (including antineutrinos) and follow a Fermi-Dirac distribution at temperature $T_{\nu,0} = 0.168$ meV~\cite{Vitagliano:2019yzm}. With energies $E_\nu \sim T_{\nu,0}$, C$\nu$B neutrinos naturally satisfy the macroscopic coherence condition $|\mathbf{q}|r_T \ll 1$ for typical interferometer targets with $r_T \sim \mu$m, making them the ideal probe of Regime III interactions.

In the cosmic microwave background (CMB) rest frame, the C$\nu$B is homogeneous and isotropic. Under this approximation, the differential flux of C$\nu$B neutrinos at Earth is derived from their number density:
\begin{eqnarray}
  n_\nu &=& \frac{1}{(2 \pi)^3}\int d^3p_\nu \,
  f_\nu(|\mathbf{p_\nu}|) =\frac{1}{(2 \pi)^3}\int 
  d\cos\theta_\nu d\phi_\nu d|\mathbf{p_\nu}| \, |\mathbf{p_\nu}|^2\,
  f_\nu(|\mathbf{p_\nu}|) \nonumber\\
  &=&\frac{1}{(2 \pi)^3}\int
    d\cos\theta_\nu d\phi_\nu dE_\nu \, |\mathbf{p_\nu}| E_\nu
\,f_\nu(|\mathbf{p_\nu}|),
\label{eq:number_density_integral}
\end{eqnarray}
where $f_\nu(|\mathbf{p_\nu}|) = [\exp(|\mathbf{p_\nu}|/T_{\nu,0}) + 1]^{-1}$ is the Fermi-Dirac distribution, $\theta_\nu$ and $\phi_\nu$ are the polar and azimuthal angles of the cosmic neutrino in the frame of the C$\nu$B rest frame, simultaneously. This yields the differential number density:
\begin{equation}
\frac{d^3 n_\nu}{d\cos\theta_\nu d\phi_\nu dE_\nu} = \frac{1}{(2 \pi)^3} \frac{E_\nu |\mathbf{p_\nu }|}{\exp(|\mathbf{p_\nu }| /T_{\nu,0}) + 1}.
\label{eq:diff_number_density}
\end{equation}

The differential neutrino flux at the detector is then:
\begin{equation}
  \frac{d^3 \Phi_\nu}{d\cos\theta_\nu d\phi_\nu dE_\nu} =
  \frac{|\mathbf{p_\nu }|}{E_\nu} \frac{d^3 n_\nu}{d\cos\theta_\nu
    d\phi_\nu dE_\nu} = \frac{1}{(2 \pi)^3} \frac{|\mathbf{p_\nu
    }|^2}{\exp(|\mathbf{p_\nu}|/T_{\nu,0}) + 1}.
\label{eq:cnub_flux_basic}
\end{equation}

Earth's motion relative to the CMB rest frame ($\beta_\oplus \sim 10^{-3}$) introduces a small but measurable anisotropy that affects both phase shift calculations and decoherence estimates. The transformation between neutrino momentum in the C$\nu$B rest frame ($\mathbf{p_\nu}$ and $\theta_\nu$) and Earth frame ($\mathbf{p'_\nu}$ and $\theta_\nu'$) is:
\begin{eqnarray}
  | \mathbf{p_\nu^\prime} | &=&
  \sqrt{| \mathbf{p_\nu}|^2 \sin^2\theta_\nu
    + \gamma^2_\oplus(| \mathbf{p_\nu}
    | \cos\theta_\nu +\beta_\oplus E_\nu)^2}, \label{eq:p_transform}\\
  \cos\theta_\nu^\prime&=& \frac{\gamma_\oplus(| \mathbf{p_\nu}
    | \cos\theta_\nu +\beta_\oplus E_\nu)}
{\sqrt{| \mathbf{p_{\nu}}|^2 \sin^2\theta_\nu
    + \gamma^2_\oplus(| \mathbf{p_\nu}
    | \cos\theta_\nu +\beta_\oplus E_\nu)^2}}, \label{eq:angle_transform}
\end{eqnarray}
where $\gamma_\oplus = (1-\beta_\oplus^2)^{-1/2} \approx 1$.

Expanding to first order in $\beta_\oplus$, the phase space transformation gives:
\begin{equation}
  d\cos\theta'_\nu \,d|\mathbf{p_\nu'}|=
  d\cos\theta_\nu \,d|\mathbf{p_\nu}|
  \left(1- \beta_\oplus
  \frac{E_\nu}{|\mathbf{p_\nu}| }\cos\theta_\nu\right),
\label{eq:phase_space_transform}
\end{equation}
yielding the Earth-frame flux:
\begin{eqnarray}
\frac{d^3 \Phi_\nu}{d\cos\theta_\nu d\phi_\nu dE_\nu}
&=& \frac{1}{(2 \pi)^3} \frac{|\mathbf{p_\nu}|^2  +
  \beta_\oplus E_\nu |\mathbf{p_\nu}|\cos\theta_\nu }
{\exp(\sqrt{|\mathbf{p_\nu}|^2  + 2 \beta_\oplus E_\nu |\mathbf{p_\nu}|
    \cos\theta_\nu}/T_{\nu,0}) + 1} \nonumber \\
&\simeq&
\frac{1}{(2 \pi)^3} \frac{|\mathbf{p_\nu}|^2}
  {\exp(|\mathbf{p_\nu}|/T_{\nu,0}) + 1} \nonumber\\
  &&\times\left[1+\beta_\oplus\cos\theta_\nu
    E_\nu
    \frac{1-\exp(-|\mathbf{p_\nu}|/T_{\nu,0})(|\mathbf{p_\nu}|/T_{\nu,0} -1)} {|\mathbf{p_\nu}|(\exp(|\mathbf{p_\nu}|/T_{\nu,0}) + 1)}
    \right].
\label{eq:fluxcnbcor}
\end{eqnarray}

The dipole anisotropy encoded in the $\beta_\oplus\cos\theta_\nu$ term provides an additional experimental handle for C$\nu$B detection, as it creates a characteristic angular dependence that helps distinguish cosmic signals from isotropic backgrounds.

Because of their ultra-low energies, the neutrino mass spectrum becomes crucial for C$\nu$B calculations. We assume normal mass ordering, massless first mass eigenstate and current best-fit values from NuFit-6.0~\cite{Esteban:2024eli}: $m_1=0$, $m_2=8.6$ meV, and $m_3=50$ meV. Each massive eigenstate contributes according to:
\begin{equation}
|\mathbf{p_{\nu i}}| = \sqrt{E_\nu^2 - m_i^2},
\label{eq:momentum_massive}
\end{equation}
with sharp kinematic thresholds at $E_\nu = m_i$ that create distinctive spectral features enhancing the experimental signature.

\subsection{Solar Neutrinos}

Solar neutrinos provide access to both Regimes 2 and 3 through two complementary production mechanisms. Nuclear fusion in the solar core produces neutrinos with energies up to several MeV~\cite{Bahcall:2004yr,Serenelli:2011pr}, while thermal processes in the solar atmosphere generate lower-energy components~\cite{Vitagliano:2019yzm,Redondo:2013lna}.

The pp-chain dominates solar neutrino production, with the primary reaction $p + p \rightarrow d + e^+ + \nu_e$ contributing flux $\Phi_{pp} = 9.96 \times 10^{10}$ cm$^{-2}$s$^{-1}$~\cite{Magg:2022rxb}. Higher energy contributions come from $^8$B decay and CNO cycle processes~\cite{Borexino:2020iwl,SNO:2013yby}. Thermal neutrino production occurs through plasma processes including photon-neutrino interactions and pair annihilation~\cite{Vitagliano:2019yzm,Redondo:2013lna}, contributing flux at eV-keV energies.

The broad energy spectrum spanning keV to MeV enables sensitivity across multiple coherence regimes within a single source. Thermal components probe atomic coherence ($|\mathbf{q}|r_{\text{atom}} \sim 1$), while nuclear components access nuclear coherence ($|\mathbf{q}|r_{\text{nuc}} \sim 1$)~\cite{Aristizabal:2022vyj,Blanco:2019lrf}. Solar neutrinos are produced as $\nu_e$ (nuclear processes) and both $\nu_e$ and 
$\bar{\nu}_e$ (thermal processes). Flavor oscillations convert these to mass eigenstates 
by the time they reach Earth. For this analysis, we neglect small flavor-dependent 
corrections in the Standard Model and assume flavor-independent BSM interactions.

\subsection{Reactor Antineutrinos}

Nuclear reactors produce $\bar{\nu}_e$ through $\beta$-decay of fission products, 
primarily from $^{235}$U, $^{239}$Pu, $^{238}$U, and $^{241}$Pu. Each fission releases 
approximately 6 antineutrinos with 200 MeV total energy, yielding $\sim 1.8 \times 10^{20}$ $\bar{\nu}_e$ 
per GW$_{\rm th}$ per second~\cite{Frampton:1982qi}.

For our calculations, we position the interferometer at 100 m from a reactor core and use 
the antineutrino spectrum compiled in Ref.~\cite{Vitagliano:2019yzm}. Additional contributions 
below the inverse $\beta$-decay threshold (1.8 MeV) come from neutron capture processes.

These three neutrino sources span nearly eight orders of magnitude in energy, from meV-scale 
C$\nu$B neutrinos to MeV-scale reactor and solar neutrinos. This broad coverage enables 
comprehensive testing of neutrino interaction models across different energy regimes.

\section{Cross Sections and Scattering Kinematics} \label{sec:CS_KIN}

Calculating neutrino-induced decoherence rates requires a comprehensive understanding of the scattering kinematics and cross sections for elastic neutrino interactions with composite targets. We consider the general elastic scattering process:
\begin{equation} 
\nu(p_\nu) + n_{\rm tgt}(p_n) \rightarrow \nu(p_\nu') + n_{\rm tgt}(q_n) 
\end{equation}
We work in the laboratory frame where the interferometer is at rest with baseline along the $\hat{z}$-axis: $\boldsymbol{\Delta x} = |\boldsymbol{\Delta x}| \hat{z}$. The target is initially at rest with four-momentum $p_n = (M, \mathbf{0})$, where $M$ represents the effective target mass that depends on the specific interaction regime.

The complete calculation of the cross sections is presented in Appendix~\ref{app:xsection}. There we derive the differential cross sections for both SM interactions and beyond-SM contributions. 

For the BSM contributions, we consider the scattering involving light vector mediators. For interactions mediated by a light neutral vector boson $Z'$ with mass $M_{Z'}$ and coupling strength $g_{Z'}$, the BSM Lagrangian is:
\begin{equation}
\mathcal{L}_V = - \sum_{i=\nu,n,p,e} g_i \overline{\psi}_i \gamma^\mu \psi_i Z'_\mu + \frac{1}{2} M_{Z'}^2 Z'^\mu Z'_\mu,
\label{eq:lagvec}
\end{equation}
where $g_i$ represents the coupling of the $Z'$ to fermion $i$.

As a representative BSM scenario, we consider the neutrinophilic $Z'$ model, where couplings to neutrinos are $\mathcal{O}(1)$ while couplings to nucleons are suppressed.
This configuration evades the stringent constraints from gravitational fifth force searches and equivalence principle tests, which apply primarily to the couplings of new mediators to nucleons and electrons, making it one of the least constrained scenarios for light mediator models.

\section{Decoherence rate} \label{sec:decoh_phase} 

After establishing the fluxes and the interactions, 
the last piece of formalism to be developed is the 
calculation of the decoherence rate 
for each type of flux per target.
Thus, the decoherence rate calculation requires careful treatment of the angular dependence of neutrino fluxes.

\subsubsection{Isotropic Sources (Cosmic Neutrino Background)}

For isotropic C$\nu$B neutrinos, we integrate over all arrival directions. Using variable transformation from $(\theta_\nu, \phi_\nu)$ to $(\gamma, \alpha)$ where:
\begin{equation}
\cos\theta_\nu = \cos\theta_q\cos\gamma + \cos\alpha\sin\theta_q\sin\gamma,
\end{equation}
with Jacobian\footnote{Details in Appendix~\ref{app:jacobian}} $d\cos\theta_\nu d\phi_\nu = d\cos\gamma d\alpha$\footnote{$\alpha$ is defined as the azimuthal relative angle between the incident neutrino momentum $\mathbf{p_\nu}$ and the momentum transfer $\mathbf{q}$. }, neglecting the anisotropy introduced by the movement of the
Earth in the C$\nu$B background, the angle $\theta_q$ only appears in the
decoherence factor which can be integrated to be

\begin{equation}
  \int_{-1}^1 d\cos\theta_q {\rm Re}[p_{\rm decoh}]= 2
  \left[ 1-\frac{\sin(|\mathbf{\Delta x}||\mathbf{q}|)} {|\mathbf{\Delta x}||\mathbf{q}|}
 \right]
\end{equation}
So in this case
\begin{eqnarray}
\int \dfrac{d \Phi_\nu}{d E_\nu} 
\,d \sigma_{\nu, 0}(M,p_{\nu}) \; {\rm Re}[p_{decoh}] \; dE_\nu
&=&
\int dE_\nu \dfrac{d \Phi_\nu}{d E_\nu}   \int d\cos\gamma 
\left[1-\frac{\sin(|\mathbf{\Delta x}||\mathbf{q})}{|\mathbf{\Delta x}||\mathbf{q}|}\right] \frac{\overline{{\cal{M}}}^\dagger \overline{\cal{M}}}{8\pi}
\dfrac{(M+E_\nu)^2 |\cos\gamma|}{\Big( (E_\nu + M)^2
  -|\mathbf{p_\nu}|^2\cos^2\gamma\Big)^2} \nonumber\\
&=&\int dE_\nu \dfrac{d \Phi_\nu}{d E_\nu} \int_{T_{\rm min}}^{T_{\rm max}} dT 
\left[1-\frac{\sin(|\mathbf{\Delta x}|\sqrt{T^2+2M T})}{|\mathbf{\Delta x}|\sqrt{T^2+2 MT}}\right]\frac{\overline{{\cal{M}}}^\dagger \overline{\cal{M}}}
     {32\pi |\mathbf{p_\nu}|^2 M}\nonumber \\
     &=&\int dE_\nu \dfrac{d \Phi_\nu}{d E_\nu}
     \int_{T_{\rm min}}^{T_{\rm max}} dT 
\left[1-\frac{\sin(|\mathbf{\Delta x}|\sqrt{T^2+2M T})}{|\mathbf{\Delta x}|\sqrt{T^2+2 MT}}\right]\frac{d\sigma_{\nu,{\rm tgt}}}{dT}     
\end{eqnarray}
where $T_{\rm max}=\frac{2 |\mathbf{p_\nu}|^2}{(M+2E_\nu+\frac{m_\nu^2}{M})}$.
The differentiating characteristic of these experiments with respect
to standard particle physics scattering probes is that, a priory,
there is no required minimum recoil kinetic energy, i.e.,  $T_{\rm min}=0$
(or equivalently to  zero value of the scattering angle $\gamma$)
which in turn implies that the calculation of the rate extends to the minimum value 
of the neutrino energy 

For the decoherence phase the relevant integral of the decoherence factor
taking into account the correction due to the $\beta_\oplus$ in Eq.~\eqref{eq:fluxcnbcor} 
\begin{eqnarray}
  \int_{_1}^1d\cos\theta_q \int_0^{2\pi}d\alpha \cos\theta_\nu {\rm Im}[p_{\rm decoh}] &=&
\int_{_1}^1d\cos\theta_q \int_0^{2\pi}d\alpha (\cos \theta_q \cos \gamma
+  \cos \alpha \sin \theta_q \sin \gamma) \sin(|\mathbf{\Delta x}||\mathbf{q}|
\cos\theta_q)\nonumber \\
&=& 4\pi\,\cos\gamma\, \left[\frac{\sin(|\mathbf{\Delta x}||\mathbf{q}|)}{(|\mathbf{\Delta x}||\mathbf{q}|)^2}-  \frac{\cos(|\mathbf{\Delta x}||\mathbf{q}|)}{|\mathbf{\Delta x}||\mathbf{q}|}
\right]
\end{eqnarray}
So in this case
\begin{eqnarray}
\int \dfrac{d \Phi_\nu}{d E_\nu} 
\,d \sigma_{\nu, 0}(M,p_{\nu}) \; {\rm Im}[p_{decoh}] \; dE_\nu
&=&
\beta_\oplus \, \int dE_\nu \dfrac{d \Phi_\nu}{d E_\nu}
    E_\nu
    \frac{1-\mathrm{exp}(-|\mathbf{p_\nu}|/{\rm T}_{\nu,0})(|\mathbf{p_\nu}|/{\rm T}_{\nu,0} -1)} {|\mathbf{p_\nu}|(\mathrm{exp}(|\mathbf{p_\nu}|/{\rm T}_{\nu,0}) + 1)} \nonumber\\
&&\hspace*{-4cm}\times \int_{T_{\rm min}}^{T_{\rm max}} dT \, \frac{(E+M)T}{\sqrt{T^2+2MT}}
\left[\frac{\sin(|\mathbf{\Delta x}|\sqrt{T^2+2M T})}{(|\mathbf{\Delta x}|\sqrt{T^2+2 MT})^2}-
\frac{\cos(|\mathbf{\Delta x}|\sqrt{T^2+2M T})}{|\mathbf{\Delta x}|\sqrt{T^2+2 MT}}
  \right]\frac{d\sigma_{\nu,{\rm tgt}}}{dT}     
\end{eqnarray}
    
\subsubsection{ Directional neutrino fluxes} 

This is the case for solar neutrinos or
  neutrinos from a reactor experiment for which the incoming neutrino
  angle is constant during the time of the measurement. There is a link
  between the angle relevant for the decoherence $\theta_q$ factor and the
  scattering angle $\gamma$  (Eq.~\eqref{eq:gamma}). The rate is
  largest when the neutrino arrives parallel to the separation
  direction  ($\cos\theta_\nu=1$). In what follows we will assume this case.
  For reactor neutrinos, this can be a realistic setup with the interferometer
  oriented in the direction of the reactor. For solar neutrinos, it would require
  the interferometer to be reoriented with the sun's position at each measurement.
  Or alternatively, one should average over the $\nu$ arrival direction, which will  
  suppress the rate by a factor ${\cal O}(10)$.
  
In this case $\gamma=\theta_q$ and
\begin{eqnarray}
\int \dfrac{d \Phi_\nu}{d E_\nu} 
\,d \sigma_{\nu, 0}(M,p_{\nu}) \; {\rm Re}[p_{decoh}] \; dE_\nu
&=&
\int dE_\nu \dfrac{d \Phi_\nu}{d E_\nu}   \int d\cos\gamma 
\left[1-\cos(|\mathbf{\Delta x}||\mathbf{q}|\cos\gamma)\right] \frac{\overline{{\cal{M}}}^\dagger \overline{\cal{M}}}{8\pi}
\dfrac{(M+E_\nu)^2 |\cos\gamma|}{\Big( (E_\nu + M)^2
  -|\mathbf{p_\nu}|^2\cos^2\gamma\Big)^2}  \nonumber\\
&
\simeq &\int dE_\nu \dfrac{d \Phi_\nu}{d E_\nu} \int_{T_{\rm min}}^{T_{\rm max}} dT
\left[1-\cos\left(\frac{|\mathbf{\Delta x}|T M }{E_\nu}\right)\right]
\frac{\overline{{\cal{M}}}^\dagger \overline{\cal{M}}}
     {32\pi E^2_\nu M}\nonumber \\
&= &\int dE_\nu \dfrac{d \Phi_\nu}{d E_\nu} \int_{T_{\rm min}}^{T_{\rm max}} dT
\left[1-\cos\left(\frac{|\mathbf{\Delta x}|T M }{E_\nu}\right)\right]
 \frac{d\sigma_{\nu,{\rm tgt}}}{dT}     
\end{eqnarray}

where in the second equality we have assumed the target mass to be much larger than
the incident neutrino energy and we have neglected the neutrino mass.
Equivalently, one finds that for the phase
\begin{equation}
\int \dfrac{d \Phi_\nu}{d E_\nu} 
\,d \sigma_{\nu, 0}(M,p_{\nu}) \; {\rm Im}[p_{decoh}] \; dE_\nu
\simeq - \int dE_\nu \dfrac{d \Phi_\nu}{d E_\nu} \int_{T_{\rm min}}^{T_{\rm max}} dT
\sin\left(\frac{|\mathbf{\Delta x}|T M }{E_\nu}\right)
\frac{d\sigma_{\nu,{\rm tgt}}}{dT}
\end{equation}

\section{Matter-Wave Interferometry Experiments}

We analyze neutrino backgrounds for two representative matter-wave interferometry experiments: 
the space-based MAQRO mission and the terrestrial Pino experiment. These experiments probe different 
regimes of the parameter space through their distinct target compositions, baselines, and operational environments.

Table~\ref{tab:mission_parameters} lists the experimental parameters used in our calculations. 
The sensitivity to neutrino interactions depends on the interferometer baseline $\Delta x$, target mass, 
measurement time $t_{\rm exp}$, and phase resolution $\sigma_\phi$.

\subsection{MAQRO}

Macroscopic Quantum Resonators (MAQRO) is a proposed space mission to
perform interferometry with high-mass objects. The mission aims to
test quantum superposition at unprecedented mass scales using SiO$_2$
nanoparticles with $10^{10}$ nucleons and a radius of 120 nm~\cite{Kaltenbaek:2015kha,Bateman:2014ena}. The
interferometer baseline separation is $\Delta x = 100$ nm, and the measurement time
is $t_{\rm exp} = 100$ s per drop. The space environment provides vacuum conditions of $\sim 10^{-17}$ mbar and eliminates 
atmospheric scattering. The solid nanoparticle configuration allows one phase measurement per drop, 
giving $\sigma_\phi = 1$ rad per measurement.

\subsection{Pino}

The ``Pino" experiment is a proposed terrestrial experiment that
utilizes an all-magnetic scheme to perform a double-slit experiment
with a macroscopic niobium sphere~\cite{Pino:reference}. This tabletop setup aims to explore
the decoherence effects from self-gravity. We assume a sphere radius
of 1 micron with $2 \times 10^{13}$ nucleons. The slit separation is $\Delta x = 290$ nm with 
free-fall time $t_{\rm exp} = 0.483$ s.. 
The experiment operates at cryogenic temperatures where niobium becomes superconducting,
 enabling magnetic control with minimal heating. 
 We assume phase sensitivity $\sigma_\phi = 1$ rad, similar to MAQRO.

\begin{table}[h!]
\centering
\caption{Experimental parameters for matter-wave interferometry experiments analyzed in this work.}
\begin{tabular}{|l|c|c|c|c|c|c|}
\hline 
Exp & Tgt & $r_{\rm tgt}$[m] & $N_{\rm nuc}$ & $\Delta x$[m] & $t_{\rm exp}$[s] & $\sigma_\phi$[rad] \\ 
\hline
MAQRO & SiO$_2$ & $1.2 \times 10^{-7}$ & $10^{10}$ & $10^{-7}$ & 100 & 1.0 \\ 
\hline 
Pino & Nb & $10^{-6}$ & $2.2 \times 10^{13}$ & $2.9 \times 10^{-7}$ & 0.483 & 1.0 \\ 
\hline
\end{tabular}
\label{tab:mission_parameters}
\end{table}

Following \cite{Du:2022ceh}, we say that an atom
interferometer has sensitivity to a given neutrino interaction
when the estimated signal is larger than the expected noise (i.e., we set a
signal-to-noise threshold of 1)
\begin{eqnarray}
    \frac{(X-X_{\mathrm{bkg}})^2}{(\sigma^T_X)^2}=1
\end{eqnarray}
where $X$ denotes either the visibility $V\equiv{\rm exp}(-s)$,
the contrast $s$ , or the
phase, $\phi$. $X_{\mathrm{bkg}}$ is the average value of either observable
without any neutrino effects, while $\sigma^T_X$ is the noise for
each observable over the full running time which we will assume to
be one year $t_{\rm tot}=1$ yr.

In what follows we will assume the ideal case with  $X_{\mathrm{bkg}}=0$. 
Furthermore  we will assume as Ref.~\cite{Du:2022ceh}
for all three
observables $\sigma^T_X$ scales with the number of measurements as
\begin{eqnarray}
  \sigma^T_X&=&\frac{\sigma_X}{\sqrt{N_{\rm meas}}}=\sigma_X
  \sqrt{\frac{t_{\mathrm{exp}}}{t_{\rm tot}}}=1.2\times 10^{-3}\;(1.2 \times 10^{-4})\nonumber\\
  &&\;\;{\rm per\; \sqrt{yr}\; at\; MAQRO\; (Pino)}
\end{eqnarray}
where $\sigma_X$ is the noise for each observable per measurement.
Furthermore we will use the values in Ref.~\cite{Du:2022ceh} for the matter
interferometers here considered $\sigma_V/V=1$ so  $\sigma_s=1$ rad
(and as mentioned above $\sigma_\phi=1$ rad as well).

\section{Results}

We present our calculations for neutrino-induced decoherence in matter-wave interferometry, considering both Standard Model interactions and beyond-Standard Model scenarios. Our results establish that while SM neutrino backgrounds are negligible, these experiments achieve competitive sensitivity to new physics.

\subsection{Standard Model Predictions}

Standard Model neutrino interactions proceed through neutral current processes mediated by $Z$ boson exchange. For the low-energy neutrino fluxes considered, the effective four-fermion interaction is characterized by vector couplings $g_V^p = 0.04$, $g_V^n = -0.5$, and $g_V^e = -0.04$.

The resulting decoherence rates are extremely small for all neutrino sources and experiments considered. Table~\ref{tab:sm_results} summarizes our findings for the signal-to-noise ratios $s/\sigma_s^T$ after one year of operation.

\begin{table}[h!]
\centering
\caption{Standard Model signal-to-noise ratios for decoherence measurements. All values are well below the detection threshold of unity.}
\begin{tabular}{|l|c|c|}
\hline
Neutrino Source & Pino & MAQRO \\
\hline
Cosmic neutrino background & $1.3 \times 10^{-22}$ & $3.7 \times 10^{-27}$ \\
Solar neutrinos & $6.4 \times 10^{-17}$ & $2.2 \times 10^{-19}$ \\
Reactor neutrinos (100 m) & $5.0 \times 10^{-15}$ & $1.7 \times 10^{-17}$ \\
\hline
\end{tabular}
\label{tab:sm_results}
\end{table}

These results confirm that Standard Model neutrino interactions produce negligible backgrounds for matter-wave interferometry experiments, simplifying the interpretation of any observed signals as evidence for new physics.

\subsection{BSM Sensitivity}

\begin{figure}
\centering
\includegraphics[width=0.6\textwidth]{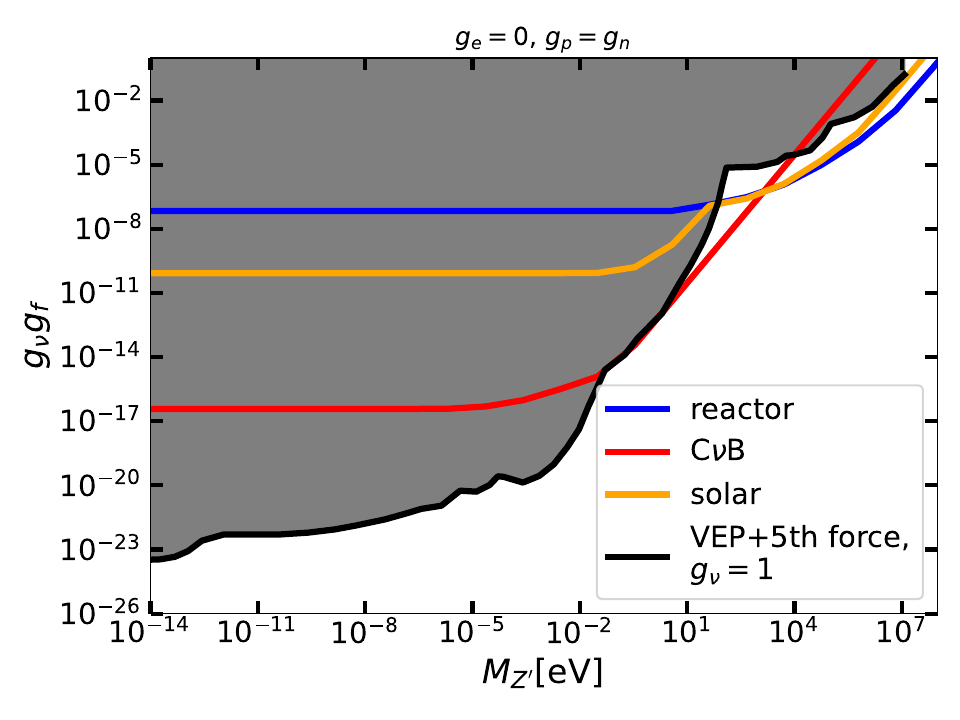}
\caption{Region in the model parameter space ($g_\nu g_n$ vs $M_{Z'}$) for which $s/\sigma^T_s=1$ in the Pino interferometer for a light $Z'$ model. The shaded region shows the parameter space where the decoherence signal would be detectable. For comparison, existing constraints from fifth force searches and equivalence principle tests are also shown, assuming a neutrinophilic $Z'$ scenario.}
\label{fig:zp}
\end{figure}

As an illustration, we show in FIG.~\ref{fig:zp} the region in the model parameter space ($g_\nu g_n$ vs $M_{Z'}$) for which $s/\sigma^T_s=1$ in the Pino interferometer for a model in which the $Z'$ couples to nucleons and neutrinos with possibly different strengths. For the sake of comparison, we also show in the figure the strongest constraints on these parameters implied by the combination of bounds from gravitational fifth force searches~\cite{Salumbides:2013dua,Adelberger:2009zz} and equivalence principle tests~\cite{Schlamminger:2007ht}. Strictly speaking, those bounds only apply to the couplings of the $Z'$ to nucleons and electrons, and some assumption must be made regarding their relation to the coupling to neutrinos. The constraints shown correspond to the least constraining assumption of a neutrinophilic $Z'$, for which the couplings to neutrinos are $\mathcal{O}(1)$ while the coupling to nucleons is suppressed. As seen in the FIG.~\ref{fig:zp}, in this case the sensitivity of the Pino interferometer is comparable to the current bounds.

The sensitivity varies depending on the neutrino source used. For the cosmic neutrino background, the interferometer achieves sensitivity to coupling products $g_\nu g_n \lesssim 10^{-17}$ for $M_{Z'} \lesssim 1$ eV, representing very competitive laboratory constraints in this mass range. Solar neutrinos probe intermediate masses with $g_\nu g_n \lesssim 10^{-11}$ for $M_{Z'} \sim$ keV, complementing other neutrino experiments. Reactor neutrinos provide controllable systematic studies with sensitivity $g_\nu g_n \lesssim 10^{-8}$ for $M_{Z'} \sim 100$ keV.

\section{Conclusions and Future Prospects}

We have presented a systematic calculation of neutrino-induced decoherence in matter-wave interferometry experiments. Our analysis treats three neutrino sources—cosmic neutrino background, solar neutrinos, and reactor antineutrinos—and accounts for coherent scattering processes across nuclear, atomic, and macroscopic length scales. The calculations provide quantitative predictions for both SM backgrounds and sensitivity to BSM physics in proposed interferometry experiments.

For SM interactions, we find that neutrino-induced decoherence remains far below experimental detection thresholds. The Pino terrestrial interferometer yields signal-to-noise ratios of $s/\sigma_s^T \sim 10^{-22}$ for cosmic neutrino background interactions, $\sim 10^{-17}$ for solar neutrinos, and $\sim 10^{-15}$ for reactor neutrinos at 100 m distance. The proposed MAQRO space mission achieves $s/\sigma_s^T \sim 10^{-27}$ for C$\nu$B interactions. These values lie many orders of magnitude below unity, confirming that Standard Model neutrino processes do not constitute a limiting background for dark matter searches using matter-wave interferometry. This situation differs markedly from direct detection experiments, where coherent neutrino-nucleus scattering establishes an irreducible background that fundamentally limits sensitivity to WIMP dark matter in certain mass ranges~\cite{Billard:2013qya,OHare:2021utq}.

The absence of significant SM backgrounds enables matter-wave interferometry to probe BSM scenarios. For interactions mediated by light vector bosons $Z'$, we find that the Pino interferometer achieves sensitivity to coupling products $g_\nu g_n \lesssim 10^{-17}$ for mediator masses $M_{Z'} \lesssim 1$ eV when using cosmic neutrino background interactions. This sensitivity becomes competitive with existing constraints from gravitational fifth force searches~\cite{Adelberger:2009zz,Salumbides:2013dua} and equivalence principle tests~\cite{Schlamminger:2007ht} under the assumption of a neutrinophilic $Z'$ model where couplings to neutrinos are $\mathcal{O}(1)$ while nucleon couplings are suppressed to satisfy gravitational constraints. Solar and reactor neutrino sources probe higher mass ranges with sensitivities of $g_\nu g_n \lesssim 10^{-11}$ for $M_{Z'} \sim$ keV and $g_\nu g_n \lesssim 10^{-8}$ for $M_{Z'} \sim 100$ keV, respectively. The interpretation of these sensitivities as actual constraints depends critically on the specific BSM model considered and the relationship between neutrino and nucleon couplings.

Several experimental and theoretical developments could enhance the physics reach of neutrino-sensitive matter-wave interferometry. On the experimental side, proposed next-generation interferometers with larger targets, longer baselines, and improved phase resolution would proportionally increase sensitivity to both decoherence and phase shifts. However, practical challenges including maintaining quantum coherence over longer timescales, controlling systematic effects, and achieving sufficient measurement statistics must be addressed. Space-based missions offer advantages in eliminating atmospheric backgrounds and enabling longer free-fall times, though they introduce substantial technical complexity and cost.

The potential for direct detection of the cosmic neutrino background deserves particular attention. While SM interactions produce unobservably small signals, BSM enhancements could bring detection within reach. The characteristic spectral features at neutrino mass thresholds ($E_\nu = m_i$) and the dipole anisotropy from Earth's motion relative to the CMB rest frame provide mechanisms for distinguishing C$\nu$B signals from other sources. However, realizing C$\nu$B detection would require not only BSM enhancements but also precise control of all systematic effects and independent confirmation through multiple experimental signatures. 

Our calculations establish the theoretical foundation for interpreting matter-wave interferometry experiments as probes of neutrino physics and BSM interactions. The complete absence of SM neutrino backgrounds simplifies the interpretation of any observed signals as evidence for new physics. At the same time, null results translate directly into constraints on BSM model parameters. As matter-wave interferometry experiments transition from conceptual proposals to precision measurements, the theoretical framework developed here provides a basis for interpreting experimental results and establishing constraints on neutrino interaction models.

\section*{Acknowledgments}

The author thanks M. C. Gonzalez-Garcia for suggesting this problem and for valuable discussions on the theoretical development and presentation of the results. 
This work has been supported by the Spanish MCIN/AEI/10.13039
/501100011033
grants  PID2022-126224NB-C21  and  the “Unit of Excellence Maria de Maeztu 2025-2029” award to the ICC-UB  CEX2024-001451-M, 
and by the European Union’s Horizon 2020 research and innovation program under the Marie Sklodowska-Curie grant HORIZON-MSCA-2021-SE- 01/101086085-ASYMMETRY.

\bibliography{references}{} \bibliographystyle{unsrt}

\onecolumngrid
\appendix

\section{Cross Section Formalism}
\label{app:xsection}

Calculating neutrino-induced decoherence rates requires a comprehensive understanding of the scattering kinematics and cross sections for elastic neutrino interactions with composite targets. We consider the general elastic scattering process:
\begin{equation}
\nu(p_\nu) + n_{\rm tgt}(p_n) \rightarrow \nu(p_\nu') + n_{\rm tgt}(q_n)
\end{equation}

We work in the laboratory frame where the interferometer is at rest with baseline along the $\hat{z}$-axis: $\boldsymbol{\Delta x} = |\boldsymbol{\Delta x}| \hat{z}$. The target is initially at rest with four-momentum $p_n = (M, \mathbf{0})$, where $M$ represents the effective target mass that depends on the specific interaction regime.

\subsection{General Cross Section Formulation}

In the laboratory frame, we define the four-momenta as:
\begin{align}
p_\nu &= (E_\nu, \mathbf{p}_\nu) = (\sqrt{|\mathbf{p}_\nu|^2 + m_\nu^2}, |\mathbf{p}_\nu|\sin\theta_\nu\cos\phi_\nu, |\mathbf{p}_\nu|\sin\theta_\nu\sin\phi_\nu, |\mathbf{p}_\nu|\cos\theta_\nu), \\
p_n &= (M, \mathbf{0}), \\
q_n &= (E_q, \mathbf{q}) = (\sqrt{|\mathbf{q}|^2 + M^2}, |\mathbf{q}|\sin\theta_q\cos\phi_q, |\mathbf{q}|\sin\theta_q\sin\phi_q, |\mathbf{q}|\cos\theta_q), \\
p_\nu' &= (E_\nu', \mathbf{p}_\nu') = (\sqrt{|\mathbf{p}_\nu'|^2 + m_\nu^2}, \mathbf{p}_\nu').
\end{align}

The differential cross section for a process with spin-averaged amplitude $\overline{|\mathcal{M}|^2}$ is:
\begin{align}
d\sigma_{\nu,\rm tgt} &= \frac{\overline{|\mathcal{M}|^2}}{2M \cdot 2|\mathbf{p}_\nu|} \frac{d^3p_\nu'}{(2\pi)^3 \cdot 2E_\nu'} \frac{d^3q}{(2\pi)^3 \cdot 2E_q} (2\pi)^4\delta^{(4)}(p_\nu + p_n - p_\nu' - q_n) \\
&= \frac{\overline{|\mathcal{M}|^2}}{16\pi^2} \frac{(M + E_\nu)^2|\cos\gamma|}{[(E_\nu + M)^2 - |\mathbf{p}_\nu|^2\cos^2\gamma]^2} d\cos\theta_q d\phi_q,
\end{align}
where $\gamma$ is the angle between the incident neutrino momentum $\mathbf{p}_\nu$ and the momentum transfer $\mathbf{q}$:
\begin{equation}
\cos\gamma = \hat{\mathbf{q}} \cdot \hat{\mathbf{p}}_\nu = \cos\theta_q\cos\theta_\nu + \cos(\phi_q - \phi_\nu)\sin\theta_q\sin\theta_\nu.
\label{eq:gamma}
\end{equation}

The momentum transfer magnitude is determined by energy-momentum conservation:
\begin{equation}
|\mathbf{q}| = \frac{2|\mathbf{p}_\nu|M(E_\nu + M)\cos\gamma}{(E_\nu + M)^2 - |\mathbf{p}_\nu|^2\cos^2\gamma} = \sqrt{T^2 + 2TM} \simeq 2|\mathbf{p}_\nu||\cos\gamma|,
\end{equation}
where the last approximation holds when $M \gg E_\nu$. The target recoil energy is:
\begin{equation}
T = E_q - M = \frac{2M|\mathbf{p}_\nu|^2\cos^2\gamma}{(E_\nu + M)^2 - |\mathbf{p}_\nu|^2\cos^2\gamma} \simeq \frac{2|\mathbf{p}_\nu|^2\cos^2\gamma}{M}.
\label{eq:recoil_energy}
\end{equation}

\subsection{Standard Model Cross Sections}

The Standard Model contributes through neutral current interactions mediated by $Z$ boson exchange. At low energies ($E_\nu \ll M_Z$), this reduces to the effective four-fermion interaction:
\begin{equation}
\mathcal{L}_{\rm SM} = \frac{G_F}{\sqrt{2}} \sum_{i=e,p,n} g_\nu \left[\overline{\psi}_\nu \gamma^\mu (1-\gamma^5) \psi_\nu\right] \left[\overline{\psi}_i \gamma^\mu (g_V^i + \gamma^5 g_A^i) \psi_i\right],
\end{equation}
with coupling constants:
\begin{align}
g_\nu &= 1, \quad g_V^p = \frac{1}{2} - 2\sin^2\theta_W = 0.0368, \\
g_V^n &= -\frac{1}{2} = -0.5, \quad g_V^e = -\frac{1}{2} + 2\sin^2\theta_W = -0.0368, \\
g_A^p &= \frac{1}{2} = -g_A^e = -g_A^n,
\end{align}
using $\sin^2\theta_W = 0.2312$.

For coherent scattering with unpolarized targets, only vector couplings contribute:
\begin{equation}
\frac{d\sigma_{\nu,\rm tgt}}{dT} = \frac{G_F^2 M}{\pi} (g_V^{\rm eff})^2 \left[1 - \frac{MT}{2E_\nu^2}\left(1 + \frac{m_\nu^2}{M^2}\right) + \frac{T}{2E_\nu}\left(\frac{T}{E_\nu} - 2\right)\right],
\end{equation}
where $g_V^{\rm eff}$ is the effective vector coupling for each interaction regime defined in Eqs.~\eqref{eq:sigma_incoh} - \eqref{eq:sigma_macro}.

\subsection{Beyond-Standard Model Cross Sections}

We consider the BSM scenario involving light vector mediators. For interactions mediated by a light neutral vector boson $Z'$ with mass $M_{Z'}$ and coupling strength $g_{Z'}$, the BSM Lagrangian is:
\begin{equation}
\mathcal{L}_V = - \sum_{i=\nu,n,p,e} g_i \overline{\psi}_i \gamma^\mu \psi_i Z'_\mu + \frac{1}{2} M_{Z'}^2 Z'^\mu Z'_\mu,
\label{eq:lagvec}
\end{equation}
where $g_i$ represents the coupling of the $Z'$ to fermion $i$.

The total squared amplitude, $\mathcal{M}^\dagger \mathcal{M}$, consists of contributions from the Standard Model and BSM processes:
\begin{equation}
\mathcal{M}^\dagger \mathcal{M} = \mathcal{M}_{\mathrm{SM}}^\dagger \mathcal{M}_{\mathrm{SM}} + 2\mathrm{Re}(\mathcal{M}_{\mathrm{SM}}^\dagger \mathcal{M}_{\mathrm{BSM}}) + \mathcal{M}_{\mathrm{BSM}}^\dagger \mathcal{M}_{\mathrm{BSM}}.
\end{equation}
Given the smallness of the SM amplitudes, the interference terms between SM and BSM amplitudes are often negligible, so we consider only the contribution from the last term. Notice that in consistency with the notation of previous equations, this amplitude is defined per $g_{e,n,p}=1$.
The spin-averaged amplitude squared is:
\begin{equation}
\overline{|\mathcal{M}_{\rm BSM}|^2} = \frac{32 g_\nu^2 M^2 E_\nu^2}{(2MT + M_{Z'}^2)^2} \left[1 - \frac{MT}{2E_\nu^2}\left(1 + \frac{m_\nu^2}{M^2}\right) + \frac{T}{2E_\nu}\left(\frac{T}{E_\nu} - 2\right)\right].
\end{equation}

This yields the enhanced cross section:
\begin{equation}
\frac{d\sigma_{\nu,{\rm tgt}}}{dT} = \frac{g_\nu^2}{2\pi} \frac{M}{(2MT + M_{Z'}^2)^2} \left[1 - \frac{MT}{2E_\nu^2}\left(1 + \frac{m_\nu^2}{M^2}\right) + \frac{T}{2E_\nu}\left(\frac{T}{E_\nu} - 2\right)\right],
\end{equation}
which can be orders of magnitude larger than SM predictions for light mediators with $M_{Z'} \lesssim$ GeV.

One example of an anomaly-free model with light mediators is the neutrinophilic $Z'$, for which the couplings to neutrinos are $\mathcal{O}(1)$ while the couplings to nucleons are suppressed. Strictly speaking, the constraints from gravitational fifth force searches and equivalence principle tests only apply to the couplings of the $Z'$ to nucleons and electrons, and some assumption must be made regarding their relation to the coupling to neutrinos. The least constraining assumption corresponds to a neutrinophilic $Z'$ model.

\section{Jacobian}
\label{app:jacobian}

The rate $R$ can be described as

\begin{eqnarray}
\int d\Omega_\nu d\Omega_q f(\gamma)
\end{eqnarray}
such that $\gamma$ is the relative angle between $\mathbf{p}_\nu$ and $\mathbf{q}$, $d \Omega_\nu= \sin \theta_\nu d \theta_\nu d \phi_\nu$
and $d \Omega_q= \sin \theta_q d \theta_q d \phi_q$ are the solid angles for  $\mathbf{p}_\nu$ and $\mathbf{q}$.
The relation between $\gamma$ and $\mathbf{p}_\nu$ and $\mathbf{q}$ can be written as:

\begin{eqnarray}
\cos \gamma = \cos \theta_\nu \cos \theta_q +  \cos (\phi_\nu-\phi_q) \sin \theta_\nu \sin \theta_q
\end{eqnarray}

Now, the objective of this section is to proof the relation

\begin{eqnarray}
\int d\Omega_\nu d\Omega_q f(\gamma) = c \int d\cos \gamma  f(\gamma)
\end{eqnarray}
such that $c$ is a constant.

In order to change variables, we must find the ideal way to transform between the old variables to the new ones.
Here, we are going to fix the old variables $\theta_q$ and $\phi_q$ and transform $\theta_\nu$ and $\phi_\nu$ into
$\gamma$ and $\beta$, such that $\theta_\nu$ is the relative angle between \{$\gamma$,$\beta$\} and \{$\theta_q$,$\phi_q$\}

\begin{eqnarray}
\cos \theta_\nu = \cos \theta_q \cos \gamma +  \cos \beta \sin \theta_q \sin \gamma
\end{eqnarray}
from this relation, it is possible to obtain 

\begin{eqnarray}
\sin \beta = \frac{\sin \theta_\nu \sin(\phi_\nu-\phi_q)}{\sin \gamma}
\end{eqnarray}

The relation between new and old variables are given by:

\begin{eqnarray}
\phi_q &=&\phi_q \nonumber\\
\theta_q &=&\theta_q \nonumber\\
\cos \gamma &=& \cos \theta_\nu \cos \theta_q +  \cos (\phi_\nu-\phi_q) \sin \theta_\nu \sin \theta_q \nonumber\\
\sin \beta &=& \frac{\sin \theta_\nu \sin(\phi_\nu-\phi_q)}{\sin \gamma}
\end{eqnarray}

The inverse of the determinant of the Jacobian of this transformation is given by,

\begin{eqnarray}
\mathrm{det}J^{-1} = \frac{\partial \gamma}{\partial \theta_\nu}\frac{\partial \beta}{\partial \phi_\nu} - 
\frac{\partial \gamma}{\partial \phi_\nu}\frac{\partial \beta}{\partial \theta_\nu}
\end{eqnarray}

Now, the partial derivatives are

\begin{eqnarray}
\frac{\partial \gamma}{\partial \theta_\nu}&=&-\frac{1}{\sin \gamma}\frac{\partial \cos\gamma}{\partial \theta_\nu} \\
\frac{\partial \gamma}{\partial \phi_\nu}&=&-\frac{1}{\sin \gamma}\frac{\partial \cos\gamma}{\partial \phi_\nu}\\
\frac{\partial \beta}{\partial \theta_\nu}&=&\frac{1}{\cos \beta}\frac{\partial \sin\beta}{\partial \theta_\nu}\\
\frac{\partial \beta}{\partial \phi_\nu}&=&\frac{1}{\cos \beta}\frac{\partial \sin\beta}{\partial \phi_\nu}
\end{eqnarray}
Now, converting the derivatives only for $\gamma$
\begin{eqnarray}
\frac{1}{\cos \gamma}\frac{\partial \sin\beta}{\partial \theta_\nu}&=&\frac{ \sin(\phi_\nu-\phi_q)}{\cos \beta \sin^2\gamma}(\cos\theta_\nu\sin \gamma + \sin \theta_\nu \cot \gamma )\frac{\partial \cos\gamma}{\partial \theta_\nu}\\
\frac{1}{\cos \gamma}\frac{\partial \sin\beta}{\partial \phi_\nu}&=&\frac{ \sin\theta_\nu}{\cos \beta \sin^2\gamma}(\cos(\phi_\nu-\phi_q)\sin \gamma + \sin (\phi_\nu-\phi_q) \cot \gamma )\frac{\partial \cos\gamma}{\partial \phi_\nu}
\end{eqnarray}

Combining all terms, the determinant of the Jacobian is written as
\begin{eqnarray}
\mathrm{det}J^{-1} &=& \frac{ 1}{\cos \beta \sin^2\gamma}(\cos\theta_\nu\sin(\phi_\nu-\phi_q)\frac{\partial \cos\gamma}{\partial \phi_\nu}-\sin\theta_\nu\cos(\phi_\nu-\phi_q)\frac{\partial \cos\gamma}{\partial \theta_\nu})\nonumber\\
&=&\frac{ \sin\theta_\nu}{\cos \beta \sin^2\gamma}(\sin\theta_\nu\cos(\phi_\nu-\phi_q) \cos\theta_q-\sin\theta_q \cos\theta_\nu)
\end{eqnarray}
 and now, substituting $\cos \beta$ we finally obtain
 \begin{eqnarray}
\mathrm{det}J^{-1} =-\frac{\sin\theta_\nu}{\sin\gamma}
\end{eqnarray}
Then, the inverse of the transformation is given by
 \begin{eqnarray}
|\mathrm{det}J |=\frac{\sin\gamma}{\sin\theta_\nu}
\end{eqnarray}
Now, since $d\Omega_q$ do not change,
\begin{eqnarray}
\int d\Omega_q= 4\pi
\end{eqnarray}
 
\begin{eqnarray}
 d\Omega_\nu   =  \sin \theta_\nu d\theta_\nu  d\phi_\nu = |\mathrm{det}J | \sin \theta_\nu d \beta d\gamma = \sin\gamma d\gamma d\beta
\end{eqnarray}

and finally
\begin{eqnarray}
\int d\Omega_\nu d\Omega_q f(\gamma)= 8\pi^2 \int_0^\pi\sin\gamma f(\gamma) d\gamma 
\end{eqnarray}

\end{document}